\documentclass[aps,prl,twocolumn,superscriptaddress,showpacs]{revtex4}

\usepackage{amsmath,bm}
\usepackage{graphicx}

\begin{document}

\title{Dynamic Jahn-Teller Mechanism of Superconductivity in MgB$_2$}

\author{Young-Woo Son}
\affiliation{School of Physics, Seoul National University, Seoul
  151-742, Korea}
\author{Jaejun Yu}
\email[Email:\ ]{jyu@snu.ac.kr}
\affiliation{School of Physics, Seoul National University, Seoul
  151-742, Korea}
\affiliation{Center for Strongly Correlated Materials Research, 
Seoul National University, Seoul 151-742, Korea}  
\author{Jisoon Ihm}
\email[Email:\ ]{jihm@snu.ac.kr}
\affiliation{School of Physics, Seoul National University, Seoul
  151-742, Korea}

\date{\today }

\begin{abstract} 
  We propose a novel mechanism of superconductivity in MgB$_2$ based on         
  the dynamic electronic structure of the $p\sigma$-orbitals coupled            
  with $e_{2g}$ phonons.  A nonconventional superconducting state 
  is found to arise from electron-phonon interactions in the 
  presence of additional pairing channels made available by the
  dynamic Jahn-Teller effects. A partially broken pseudo-spin symmetry 
  in this Jahn-Teller system, together with two-phonon exchange pairing, 
  naturally gives rise to two distinct gaps both of 
  which are basically isotropic in the        
  ($k_x$, $k_y$) space.  Important experimental observations including          
  high $T_c$ and the anomalous specific heat are explained using this           
  theory.  
\end{abstract}
\pacs{74.20.Fg, 71.70.Ej, 74.25.Jb, 74.70.Ad}
%
%
%
%
%
\maketitle


Following the discovery of high $T_c$ superconductivity near 40 K in
MgB$_2$~\cite{akimi01}, a lot of efforts have been devoted to revealing
the underlying mechanism of this novel superconductor~\cite{buzea01}.
 Now it is widely
agreed upon that a relatively strong electron-phonon coupling is 
responsible for the superconductivity of this material and there
exit, at least phenomenologically, 
two superconducting gaps~\cite{shin01,chen,szabo,bouquet01a,junod01,yang01,bouquet01b}.
It has been proposed~\cite{kortus01b,choi01a} 
that the complex Fermi surfaces having 
both 2 dimensional (2D) cylindrical sheets 
and 3 dimensional (3D) tubular networks produce
two gaps. 
However, it is still an open question whether one gap is mainly
associated with the 2D cylindrical Fermi surfaces and the other
is associated with the 3D Fermi surfaces. 

In this letter, we propose a mechanism of superconductivity in MgB$_2$
based on a dynamic Jahn-Teller effect arising from the interplay
between the doubly degenerate $p\sigma$ electronic states and the
$e_{2g}$ phonon modes. The hopping motion of holes of the $p\sigma$
character on the Boron layers is constrained by the `pseudo-spin', which will be
defined later, and partial breakdown of the pseudo-spin 
symmetry is also important in the pairing mechanism.
Unlike existing theories, we will explain both 
high $T_c$ and two distinct gaps in MgB$_2$ 
without invoking the additional 3D Fermi surface of $p\pi$ character. 

According to the band structure
calculations~\cite{kortus01a,pickett01a}, the $p\sigma$ states on the
Boron layer form doubly degenerate $E_g$ bands at $\Gamma$ (to be
precise, along the $\Gamma$A line on which the component of $\mathbf{k}$ parallel
with the Boron plane, $\mathbf{k}_{||}$, is zero), where we have
small hole pockets presumably responsible for the 
superconductivity in MgB$_2$. (Splitting of the band degeneracy away from
$\Gamma$ will be taken into account later.)  These hole states
are known to be strongly coupled with $e_{2g}$ phonon modes
with $\hbar\omega\approx70meV$~\cite{cava01b,kong01}.  The $E_g$ hole
doublet coupled with $e_{2g}$ phonon doublet states on the Boron layer
can be represented by the so-called $E\otimes e$ Jahn-Teller model.  A
typical adiabatic potential energy surface of the $E\otimes e$
Jahn-Teller model is illustrated in Fig.~1, where $q_\theta$ and
$q_\varepsilon$ correspond to the two phonon coordinates.
 \begin{figure}[b]
  \centering
  \includegraphics[width=6cm]{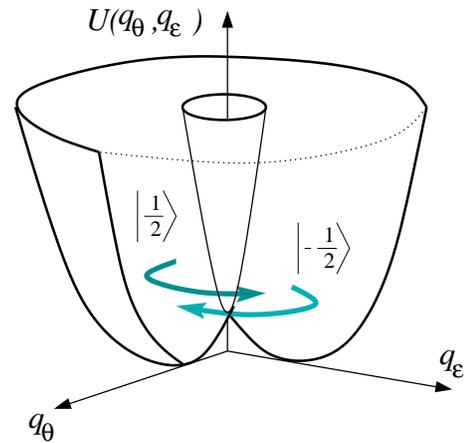}
  \caption{Adiabatic potential surface of the $E\otimes e$ Jahn-Teller system and the
           doubly degenerate pseudo-spin states
           ($|\frac{1}{2}\rangle$, $|-\frac{1}{2}\rangle$) of the hole.}
  \label{fig:1}
\end{figure}

It is well known that one must take account of the dynamical phases of
the holes and phonons in order to describe the $E\otimes e$ system
properly.  When the ionic configuration traverses around the potential
well in Fig.~1 once, the corresponding {\it orbital} wavefunction of the hole
(for simplicity of discussion, we leave aside the 
real spin wavefunction which simply follows the elementary BCS theory) 
acquires a phase of $-1$, as if the orbital part had spin
1/2. We call this a pseudo-spin of the hole and denote its 
projected value by $\alpha=\pm\frac{1}{2}$.  This subject was
extensively discussed in the literature~\cite{bersuker} in the context
of the Jahn-Teller effect or Berry's phase.  We also note that the
dynamic Jahn-Teller effect in doped C$_{60}$ superconductors was studied
before, although the methods and results there were different from our
work~\cite{auer}.

We introduce a single site Jahn-Teller Hamiltonian per unit cell
using mass-weighted coordinate for the symmetrized displacement
$Q_\gamma$($\gamma=\theta,\varepsilon$), i.e. 
$q_\gamma = \sqrt{M_\gamma}Q_\gamma$,
\begin{equation}
 h_o=\frac{1}{2}(p^2_{\theta}+p^2_{\varepsilon})
  +\frac{1}{2}\omega^2_0 (q^2_{\theta}+q^2_{\varepsilon})
  +A(q_{\theta}{\sigma}_x +q_{\varepsilon}{\sigma}_y ),
\end{equation}  
where $A$ is the hole-phonon coupling strength and $\sigma_x$,
$\sigma_y$ are Pauli spin matrices for the pseudo-spin 
representation of $E_g$ hole doublet states.  In an analogy to the
2D harmonic oscillator, the vibrational mode in the
$(q_{\theta},q_{\varepsilon})$-space belongs to the eigenmode of the
``phonon'' angular momentum operator $L_z$
(=$q_{\theta}p_{\varepsilon}-q_{\varepsilon}p_{\theta}$) with the
eigenvalue of $l$, and the principal quantum number $\nu$ 
becomes $\nu=l+2k$ ($k$ is a non-negative integer and $-\nu\le l \le\nu$).  
In the present theory of superconductivity, only $\nu=$ 0, 1, 2 and $l=0,\pm 1$
are relevant.  Defining the total pseudo-spin of the system by
$J_z=L_z+\frac{1}{2}\sigma_z$, it is straightforward to prove 
$[h_0, J_z]=0$, that is, the total pseudo-spin of the system is
conserved, while individual $L_z$ and $\sigma_z$ are not.  
(The phonon angular momentum component, $L_z$,
can be regarded as a part of the total pseudo-spin in this context.)
These pseudo-spin states are derived from the $E\otimes e$
Jahn-Teller coupling between $p\sigma$ holes and $e_{2g}$
phonons, and reflect the non-adiabatic nature of the coupled states.
As a consequence, relevant electron and phonon states are
expected to be strongly renormalized and contribute to anomalous
behaviors (e.g., broad line-widths) observed in recent photoemission spectroscopy and 
Raman experiments~\cite{shin01,takahashi,szabo}.   
\begin{figure}[t]
  \label{fig:2}
  \includegraphics[width=8cm]{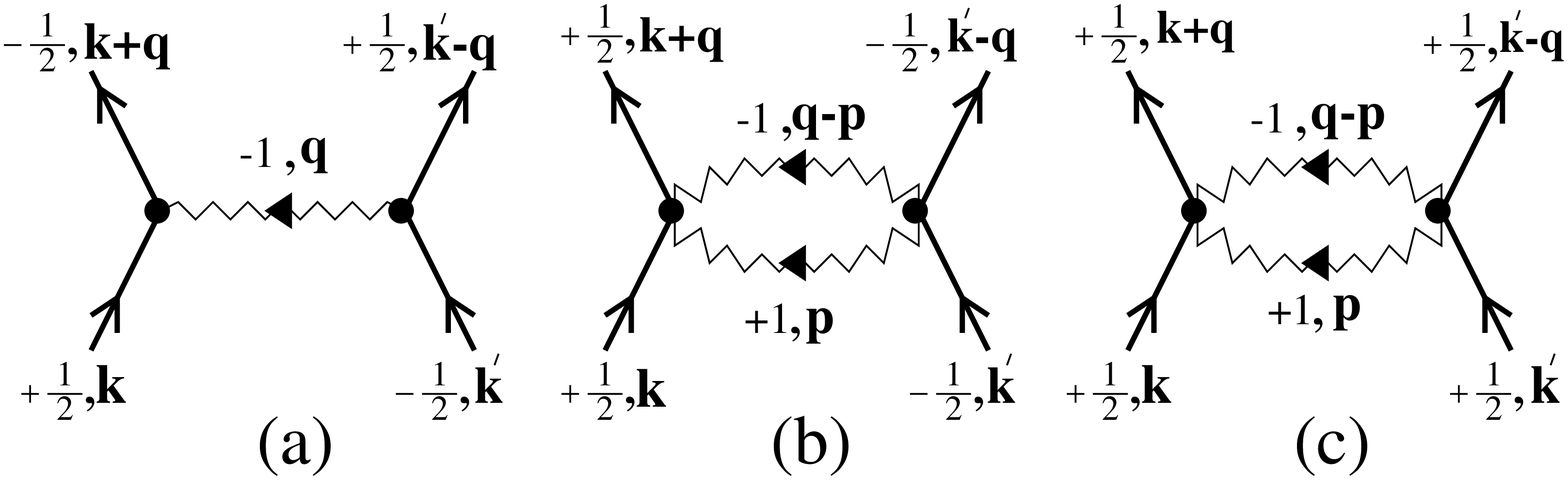}
  \caption{Diagrams for pairing interactions with phonon exchange : 
    (a) one-phonon exchange, and (b),(c) two-phonon exchange. 
     The numbers indicate pseudo-spins. The obvious real spin indices are 
     omitted for notational clarity. Diagrams with all the pseudo-spin signs 
     switched are allowed as well (not shown).}
\end{figure}

We generalize the theory to the Bloch state in the $\mathbf{k}$-space of
a crystal by linear combinations of the local holes of the same
pseudo-spin, but mathematical details will be reserved for a longer
paper~\cite{YWS}.  We first consider the one-phonon ($\nu=1$) exchange
for the Cooper pair formation. Each phonon has $l=\pm 1$ in this case.
In the one-phonon absorption or emission, the pseudo-spin of the hole
should flip to conserve the total pseudo-spin. The only allowed
one-phonon exchange mechanism is the opposite pseudo-spin Cooper pair as
illustrated in Fig.~2(a). Pairing of the same pseudo-spin holes
violates the conservation law in either one of the two (emission and
absorption) processes. Such a selection rule has already been known in
the self-energy correction to the Jahn-Teller system~\cite{takada}.

We next consider the two-phonon ($\nu=2$) case.  In MgB$_2$, the
two-phonon exchange process has been thought to be important in
producing high $T_c$ and included in some previous
works~\cite{kortus01b}.  When we properly consider the pseudo-spin
symmetry of the Jahn-Teller system such as MgB$_2$, 
incorporating the two-phonon exchange turns out to be essential in order to obtain two
distinct gaps.  Unlike the one-phonon exchange process, both opposite-
and same- pseudo-spin pairs are allowed here as illustrated in Fig.~2(b) 
and (c). The total $l$ (sum of two individual $l$'s) 
of intermediate phonons is restricted to zero
(i.e., the hole pseudo-spin change of $\pm 2$ is not
possible) to conserve the total pseudo-spin.  A new model Hamiltonian
with different pseudo-spin channels explicitly labelled is, in the
$\mathbf{k}$-space representation,
\begin{equation}
  \label{eq:12}
   \mathcal{H} = \mathcal{H}_0 + \mathcal{H}_1 + \mathcal{H}_2 + \mathcal{H}_3 . 
\end{equation}
\begin{subequations}
  \label{eq:total}
  \begin{eqnarray}
    {\cal{H}}_0 
    &=& \sum_{\alpha {\bf k} \sigma } 
    \varepsilon_{{\bf k}} c^{\dagger}_{\alpha {\bf k} \sigma } c_{\alpha
      {\bf k} \sigma }, \label{eq:12a} \\ 
    {\cal{H}}_1 
    &=& -\sum_{\alpha {\bf k} {\bf k}'}
    V^{0}_{{\bf k} {\bf k}'} 
    c^{\dagger}_{-\alpha{\bf k}'\uparrow}c^{\dagger}_{\alpha-{\bf k}'\downarrow}
    c_{-\alpha-{\bf k} \downarrow}c_{\alpha {\bf k} \uparrow},
    \label{eq:12b}\\ 
    {\cal{H}}_2 
    &=& -\sum_{\alpha {\bf k} {\bf k}'}
    V^{1}_{{\bf k} {\bf k}'} 
    c^{\dagger}_{\alpha{\bf k}'\uparrow}c^{\dagger}_{-\alpha-{\bf k}'\downarrow}
    c_{-\alpha-{\bf k} \downarrow}c_{\alpha {\bf k} \uparrow},
    \label{eq:12c}\\ 
    {\cal{H}}_3 
    &=&-\sum_{\alpha {\bf k} {\bf k}'}
    V^{2}_{{\bf k} {\bf k}'} 
    c^{\dagger}_{\alpha{\bf k}'\uparrow} c^{\dagger}_{\alpha-{\bf k}'\downarrow}
    c_{\alpha-{\bf k} \downarrow}c_{\alpha {\bf k} \uparrow}.
     \label{eq:12d} 
  \end{eqnarray}
\end{subequations}
$\mathcal{H}_1$, $\mathcal{H}_2$, and $\mathcal{H}_3$ correspond to the
diagrams in Fig. 2(a), (b), and (c), respectively.  The explicit form of
the derived interactions, $V^i_{\mathbf{k k'}}$'s ($i=$0, 1, 2), can be
calculated and will be given elsewhere~\cite{YWS}. To discuss
the nature of superconducting states, it suffices here to treat them
as given parameters.  The additional pairing channels due to the presence of
pseudo-spin symmetry can give rise
to a non-trivial superconducting ground state.  Although the
pairing is still based on electron-phonon
interactions, the pairing order parameter can now attain more than one
symmetry representation leading to the ground state with
multi-dimensional order parameters while retaining the conventional
$s$-wave type superconductivity. 

In the pairing Hamiltonian described above, however, important features
of a real crystal are missing. First, the pseudo-spin symmetry is exact
only at $\mathbf{k}_{||}=0$ and there occurs small but finite deviation
from the exact symmetry for $\mathbf{k}_{||}\neq0$ where Fermi surfaces
actually exist. The band splitting at $\mathbf{k}_{||}\neq0$ is a clear
manifestation of the breakdown of the rigorous Jahn-Teller theory for
Bloch wavefunctions of $\mathbf{k}_{||}\neq0$ in a crystal as opposed to
the case of a highly symmetric molecule.  Second,
higher-order Jahn-Teller coupling terms (not considered so far) do not
commute with $J_z$ and break the pseudo-spin conservation~\cite{pseudo}.
A simple way to incorporate the symmetry breaking effect is to introduce
a small symmetry breaking parameter $\phi_\mathbf{k}$
($\phi_{\mathbf{k}_{||}=0}=0$ and $\phi_{\mathbf{k}_{||}\neq 0}\neq 0$)
in the hopping Hamiltonian (Eq. (3a)) as follows,
\begin{equation}
\tilde{\mathcal{H}_0}
=\sum_{\alpha {\bf k} \sigma } 
 \varepsilon_{\bf k} c^{\dagger}_{\alpha {\bf k} \sigma } 
                           c_{\alpha {\bf k} \sigma }
-\sum_{\alpha {\bf k} \sigma } 
 \phi_{\bf k}\varepsilon_{\bf k} c^{\dagger}_{\alpha {\bf k} \sigma } 
                           c_{-\alpha {\bf k} \sigma }.
\end{equation} 
This replacement naturally introduces 
the symmetry-breaking hopping process which gives the band splitting
away from $\mathbf{k}_{||}=0$. 
(If the pseudo-spin $\frac{1}{2}$ were
continuous symmetry ($SU(2)$) for all $\mathbf{k}$'s, there would
be a single finite gap plus a zero-gap Goldstone mode, unlike
the experimental observation in MgB$_2$.)
Now we choose the following two order parameters, one for
the opposite pseudo-spin pair and the other for the same 
pseudo-spin pair, 
\begin{subequations}
  \begin{eqnarray}
& A_\mathbf{k}=\langle c_{-\frac{1}{2}-\mathbf{k}\downarrow}
                       c_{\frac{1}{2}\mathbf{k}\uparrow} \rangle
              =\langle c_{\frac{1}{2}-\mathbf{k}\downarrow}
                       c_{-\frac{1}{2}\mathbf{k}\uparrow} \rangle,\\
& B_\mathbf{k}=\langle c_{\frac{1}{2}-\mathbf{k}\downarrow}
                       c_{\frac{1}{2}\mathbf{k}\uparrow} \rangle
              =\langle c_{-\frac{1}{2}-\mathbf{k}\downarrow}
                       c_{-\frac{1}{2}\mathbf{k}\uparrow} \rangle.
 \end{eqnarray}
\end{subequations}
It is to be noted that the pseudo-spin pairing assumed above is 
a triplet state (even for exchange of two particles) to preserve
the singlet (odd) state of the real spin pair and the s-wave (even)
state of the orbital wavefunction.
Following the standard mean field approximation, we can derive
the effective Hamiltonian $\mathcal{H}'$
 ($=\mathcal{H}-\varepsilon_F\mathcal{N}$), 
\begin{eqnarray}
\mathcal{H}'&\simeq&\sum_{\alpha\mathbf{k}}
\xi_\mathbf{k}(c^\dagger_{\alpha\mathbf{k}\uparrow}
               c_{\alpha\mathbf{k}\uparrow}
              -c_{-\alpha-\mathbf{k}\downarrow}
               c^\dagger_{-\alpha-\mathbf{k}\downarrow})\nonumber\\
& &-\sum_{\alpha\mathbf{k}}
\delta_\mathbf{k}(c^\dagger_{\alpha\mathbf{k}\uparrow}
               c_{-\alpha\mathbf{k}\uparrow}
              -c_{-\alpha-\mathbf{k}\downarrow}
               c^\dagger_{\alpha-\mathbf{k}\downarrow})\nonumber\\
& &-\sum_{\alpha\mathbf{k}}
\Delta_{1,\mathbf{k}}(c^\dagger_{\alpha\mathbf{k}\uparrow}
               c^\dagger_{-\alpha-\mathbf{k}\downarrow}
              +c_{-\alpha-\mathbf{k}\downarrow}
               c_{\alpha\mathbf{k}\uparrow})\nonumber\\
& &-\sum_{\alpha\mathbf{k}}
\Delta_{2,\mathbf{k}}(c^\dagger_{\alpha\mathbf{k}\uparrow}
               c^\dagger_{\alpha-\mathbf{k}\downarrow}
              +c_{\alpha-\mathbf{k}\downarrow}
               c_{\alpha\mathbf{k}\uparrow})\nonumber\\
& &+\text{(constant terms)},
\end{eqnarray}
where $\xi_\mathbf{k}=\varepsilon_\mathbf{k}-\varepsilon_F$, 
$\delta_\mathbf{k}=\phi_\mathbf{k}\varepsilon_\mathbf{k}$,
and the gap-like paremeters are expressed as  
\begin{equation}                                                     
  \Delta_{1,\mathbf{k}}                                              
  =\sum_\mathbf{k'}(V^0_\mathbf{kk'}+V^1_\mathbf{kk'})A_\mathbf{k'},
  \;\;\;\;
  \Delta_{2,\mathbf{k}}                                              
  =\sum_\mathbf{k'}V^2_\mathbf{kk'}B_\mathbf{k'}.
\end{equation}
Using the basis 
$(c^\dagger_{\frac{1}{2}\mathbf{k}\uparrow},
c^\dagger_{-\frac{1}{2}\mathbf{k}\uparrow},
c_{\frac{1}{2}-\mathbf{k}\downarrow},
c_{-\frac{1}{2}-\mathbf{k}\downarrow})$, it is convenient to cast
the Hamiltonian into a matrix form (by omitting of the subscript $\mathbf{k}$
for notational simplicity),
\begin{equation}
\mathcal{H}'=\sum_\mathbf{k}
\left(
\begin{array}{cccc}
\xi&-\delta&-\Delta_2&-\Delta_1\\
-\delta&\xi&-\Delta_1&-\Delta_2\\
-\Delta_2&-\Delta_1&-\xi&\delta\\
-\Delta_1&-\Delta_2&\delta&-\xi
\end{array}
\right).
\end{equation}
This Hamiltonian is easily diagonalized, for each $\mathbf{k}$, 
yielding two positive eigenvalues,
\begin{equation}
E_\pm=\sqrt{(\xi\mp\delta)^2+(\Delta_2 \pm\Delta_1)^2}.
\end{equation}
The larger gap $\Delta_+(=\Delta_2 +\Delta_1)$ is associated with
the band of $E_{\text{normal}}=\xi-\delta$ and the
smaller gap $\Delta_-(=|\Delta_2 -\Delta_1|)$ is associated with the
band of $E_{\text{normal}} =\xi+\delta$.
Because of this band splitting, two gaps are associated with two 
different normal-state (above $T_c$) densities of states (DOSs). 
If we define $N$ to be the normal-state DOS of unsplitted $p\sigma$ band
at the Fermi level before the symmetry-breaking $\phi$ is introduced,
normal-state DOSs of two bands are  
$N_\pm =\frac{N}{1\mp\phi}$
with the 
assumption of $\phi_\mathbf{k}$'s being independent of $\mathbf{k}$
 ($\phi_\mathbf{k}=\phi$) in the vicinity of the Fermi level. 
Experimentally, it is observed that
the state involved in the larger gap has the larger normal-state DOS 
than that of the smaller one~\cite{bouquet01b}.
In this situation, we naturally assign a positive value for 
$\phi$ ($N_+ =\frac{N}{1-\phi}>N_- =\frac{N}{1+\phi}$).

We first follow, for heuristic purposes, the approximation procedure
of the weak-coupling BCS theory. We assume some {\it effective} cutoff energy
$\omega$ and define $\lambda_1=N(V^0 +V^1)$ and $\lambda_2=NV^2$. 
Only $\lambda_1$, $\lambda_2$ and $\omega$ are independent variables
for the gap equations.
The coupled gap equations for finite temperature read as
\begin{eqnarray}
\Delta_+&=&\frac{\lambda_2+\lambda_1}{1-\phi}\int^{\omega}_0
d\xi\frac{\Delta_+}{2E_+}\tanh\frac{\beta E_+}{2}\nonumber\\
& &
+\frac{\lambda_2-\lambda_1}{1+\phi}\int^{\omega}_0 
d\xi\frac{\Delta_-}{2E_-}\tanh\frac{\beta E_-}{2}~,\nonumber\\
\Delta_-&=&\frac{\lambda_2+\lambda_1}{1+\phi}\int^{\omega}_0
d\xi\frac{\Delta_-}{2E_-}\tanh\frac{\beta E_-}{2}\nonumber\\
& &
+\frac{\lambda_2-\lambda_1}{1-\phi}\int^{\omega}_0 
d\xi\frac{\Delta_+}{2E_+}\tanh\frac{\beta E_+}{2}~.
\end{eqnarray}
By fitting to $T_c$ and the specific heat behavior, we are going to
obtain below $\lambda_1 =0.235$, $\lambda_2 =0.25$ and $\phi=0.2$ for
the assumed value of $\hbar\omega=74.5meV$.  The normalized solutions of the
above gap equations (with $\frac{\lambda_1}{\lambda_2}\simeq0.94$ 
and $\phi=0.2$) are
shown in Fig. 3(a).  Our theory gives two distinct gaps of the identical
$T_c$ originated from two split-off $p\sigma$ bands alone, in contrast
to other theories which invoke the $p\pi$ band to obtain multi-gaps.
Furthermore, we predict that there exist two gaps for the same
$\mathbf{k}$ vector and both gaps are basically isotropic in the Boron
plane (except for a small anisotropic influence from the $p\pi$ bands
located at different regions of the $\mathbf{k}$ space) unlike other
theories proposing generically anisotropic gaps.    
Without pseudo-spins,
there would be only one s-wave gap solution giving the lowest free
energy.  It should be possible, with angle-resolved experimental probes,
to test the symmetry or angular dependence of the gap order
parameters. Also, it is important in the present theory to include the
symmetry-breaking parameter $\phi_\mathbf{k}$; otherwise, two gaps
should be exactly degenerate and only one gap would manifest itself.
With the same degree of approximation as in the BCS formalism, we can
derive the expression for $T_c$ (weak coupling limit),
\begin{equation}
\label{temperature}
k_B T_c\simeq1.13\hbar\omega\exp
\left[-\frac{(\lambda_1+\lambda_2)
-\sqrt{(\lambda_1-\lambda_2)^2+4\phi^2 \lambda_1 \lambda_2}}
{2\lambda_1 \lambda_2}
\right].
\end{equation}

We then calculate the specific heat with this gap structure.  The
resulting specific heat is plotted in a usual manner in Fig. 3(b).  To
obtain both $T_c =39~K$ and the specific heat data in agreement with
experiment, we require $\phi=0.2$, $\lambda_1=0.235$, $\lambda_2=0.25$.
The symmetry breaking parameter $\phi=0.2$ gives about the same band
splitting as {\it ab initio} calculations for the two $p\sigma$ 
bands near the Fermi level.  Needless to
say, since $T_c$ is high and the material parameters are not in the
weak-coupling limit, Eq.~(11) is not to be applied directly to MgB$_2$. We
have to use a more realistic (Eliashberg) equation which gives, for example,
$T_c\simeq \frac{\langle \omega_{\text{ln}} \rangle}{1.20}
\exp\left[-\frac{1.04(1+\lambda)}{\lambda-(1+0.62\lambda)\mu^*}\right]
$~\cite{allen}.  It is beyond the scope of the present paper to calculate
$\lambda_1$ and $\lambda_2$ from the first principles. The effective
coupling constant $\lambda$ in the exponent in Eq.~(\ref{temperature}),
namely, $ \frac{2\lambda_1 \lambda_2} {(\lambda_1+\lambda_2)-
  \sqrt{(\lambda_1-\lambda_2)^2+4\phi^2 \lambda_1 \lambda_2}} $ should
be about 0.95, as is well known ~\cite{pickett01a,kong01}, 
in order to give $T_c =39$~K with $\mu^*=0.13$.  
Then we obtain
$\lambda_1 =0.71$, $\lambda_2 =0.76$ by simple scaling of the previous values.  
These values are not to be taken literally because the two-phonon exchange
can modify the empirical expression for $T_c$ of the Eliashberg equation.
In any cases, two-phonon exchange together with extra pairing channels 
makes an important contribution to $T_c$ and
it is the source of larger effective $\lambda$ in MgB$_2$ than in other materials.
\begin{figure}[t]
  \label{fig:3}
   \includegraphics[width=8cm]{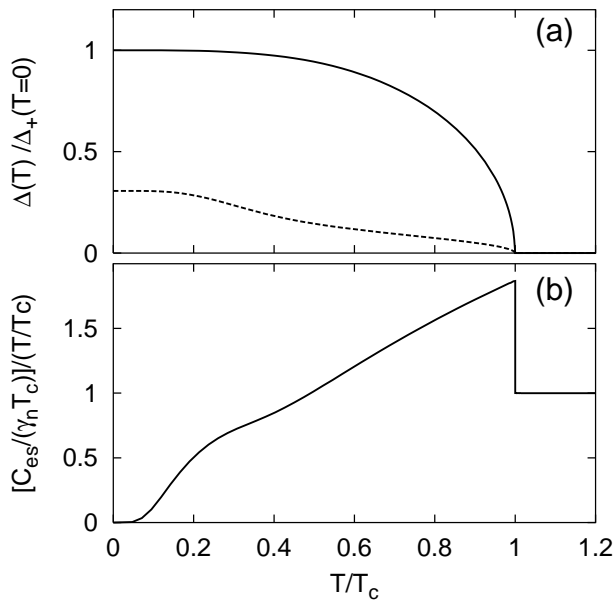}
 \caption{(a) Solution of the self-consistent equations for
         two gaps $\Delta_+$ and $\Delta_-$, and (b) calculated specific heat 
         for the above gap structure. }
\end{figure}

In summary, we have proposed a pairing mechanism in superconducting 
MgB$_2$ based on the dynamic Jahn-Teller
effect of the doubly degenerate $p\sigma$ hole states coupled to 
degenerate $e_{2g}$ phonons.  The pseudo-spin contraint
imposed on the motion of holes by the
Jahn-Teller interactions is an important factor in understanding the
dynamical properties of the system.  
The one-phonon and two-phonon couplings 
lead to effective pairing interactions with additional pseudo-spin
channels.  As results, the high $T_c$ observed in MgB$_2$ is
attributed mainly to the two-phonon exchange mechanism, and further the
symmetry breaking hopping term plays a crucial role in producing 
two gaps which are responsible for
the anomalous behaviors observed in experiment 
including the specific heat measurement.

We are grateful to Prof. T.W. Noh for helpful
discussions. This work was supported by the BK21 Project of KRF.  
J.Y. acknowledges the support by the KOSEF through CSCMR. 
Y.-W.S and J.I. acknowledge the support by the KOSEF through
the CNNC of the Sungkyunkwan University.


\begin{thebibliography}{99}
\bibitem{pseudo}To be precise, the off-diagonal second-order 
  and the diagonal third-order coupling break the symmetry.
\end{thebibliography}
\end{document}